\documentclass[preprint]{elsarticle}
\makeatletter
\def\ps@pprintTitle{%
 \let\@oddhead\@empty
 \let\@evenhead\@empty
 \def\@oddfoot{\centerline{\thepage}}%
 \let\@evenfoot\@oddfoot}
\makeatother

\usepackage{amsmath,amssymb}
\usepackage{amstext}
\usepackage{amsbsy}
\usepackage{graphics}
\usepackage{latexsym}
\usepackage{verbatim}

\def\reals{{{\rm l} \kern-.15em {\rm R}}}
\newcommand{\x}{\mathbf{x}}
\newcommand{\y}{\mathbf{y}}

\newcommand{\p}{\mathbf{p}}
\renewcommand{\k}{\mathbf{k}}
\newcommand{\kbar}{\mathbf{\bar{k}}}
\newcommand{\kappab}{\boldsymbol{\kappa}}

\newcommand{\Z}{\mathbb{Z}}
\newcommand{\R}{\mathbb{R}}

\newcommand{\erf}{\text{erf}}
\newcommand{\erfc}{\text{erfc}}

\renewcommand{\r}{\mathbf{r}}

\title{The Ewald sums for singly, doubly and triply periodic \\ electrostatic systems}
\author{Anna-Karin Tornberg\tnoteref{t1}}
\tnotetext[t1]{E-mail: \texttt{akto@kth.se}.}
\address{KTH Mathematics, Swedish e-Science Research Center, 100 44
  Stockholm, Sweden.}

\begin{document}
\bibliographystyle{plain}

\pagestyle{myheadings}
\thispagestyle{plain}
\thispagestyle{empty}

\begin{abstract}
 When evaluating the electrostatic potential, 
 periodic boundary conditions in one, two or three of the spatial
 dimensions are often needed for different applications. The triply
 periodic Ewald summation formula is classical, and Ewald summation
 formulas for the other two cases have also been derived. 
In this paper, derivations of the Ewald sums in the doubly and singly
periodic cases are presented in a uniform framework based on Fourier
analysis, which also yields a natural starting point for FFT-based
fast summation methods. 
\end{abstract}

\maketitle

\section{Introduction}

A fundamental task in electrostatics is to compute the potential due
to a number of charged particles. Periodic boundary conditions in all
spatial directions are often applied to emulate properties of a larger
aggregate. In simulations of liquid or solid surfaces and membranes
etc., it is often desirable to apply periodicity only in two of three
spatial directions, and considering geometries involving pores or
channels, one-dimensional periodic boundary conditions are useful. 

Assume that we have $N$ particles with charge $q_n$ located at $\x_n$,
$n=1,\ldots,N$, in a domain 
$\Omega=[-L_1 /2, L_1 /2] \times [-L_2 /2, L_2 /2]  \times [-L_3 /2,
L_3 /2]$, 
where the system is charge neutral, i.e. 
$\sum_{n=1}^N q_n \equiv 0$. 
The electrostatic potential due to these charges, evaluated at these
same locations, is given by the sum 
\[
\phi^{DP}(\x_m)= \sum_{\p \in P_D} \sum_{n=1}^{N,'} 
\frac{q_n}{|\x-\x_n+\p |}, \quad m=1,\ldots,N. 
\]
The sum over $\p$ is a periodic replication of the charges, and
$D=1,2,3$ indicates the number of periodic directions.  
The ${N,'}$ indicates that the term ($n=m$, $\p={\bf 0}$) is excluded
from the sum. 
We define
\begin{align} 
P_3 & =\{ (jL_1,lL_2,pL_3\}: (j,l,p) \in \Z^3 \}, \quad 
P_2  = \{ (jL_1,lL_2,0)\}: (j,l) \in \Z^2 \}, 
\notag \\
P_1 & = \{ (0,0,pL_3\}: p \in \Z \}. 
\label{eqn:perdef}
\end{align}
Here, we have chosen $x$ and $y$ as the periodic directions and $z$ as
the free direction in the doubly periodic case (2P), and $x$ and $y$
as the free and $z$ as the periodic direction in the singly periodic
case (1P). 

In the triply periodic case, the sum given above is only conditionally
convergent also for charge neutral systems, and the result will depend on the
summation order. 
Ewald \cite{Ewald1921}, showed that the
potential can be computed by splitting the contribution from each
charge into a rapidly decaying part and a
smooth part which is summed in Fourier space. 
This yields a well-defined expression that corresponds to a spherical
summation order of the original sum. 
The Ewald sum for evaluating the potential at a source location
$\x_m$, $m=1,\ldots,N$ 
under triply periodic boundary conditions is
\begin{align}
\phi^{3P}(\x_m)= & \sum_{\p \in P_3} \sum_{n=1}^{N,'} q_n \frac{
\text{erfc}(\xi \, |\x_m-\x_n+\p |)}{|\x_m-\x_n+\p |} + 
\notag \\
& +\frac{4 \pi}{V} \sum_{\k \ne {\bf 0}} \frac{e^{-k^2/4\xi^2}}{k^2}
\sum_{n=1}^N q_n e^{-i \k \cdot (\x_m-\x_n)} \ 
 - \frac{2 \xi}{\sqrt{\pi}}q_m. 
\label{eqn:Ewald3P}
\end{align}
Here, the ${N,'}$ indicates that the term ($n=m$, $\p={\bf 0}$) is excluded from the real space
sum and $P_3$ is given in (\ref{eqn:perdef}).
The $\k$-vectors form the discrete set 
$\{ 2\pi ( \frac{n_1}{L_1}, \frac{n_2}{L_2}, \frac{n_3}{L_3} ): (n_1,n_2,n_3) \in Z^3 \}$, $k^2=|\k|^2$ and $V=L_1 L_2 L_3$.
Here, $\xi>0$ is the decomposition parameter. The result is
independent of this parameter, but it controls the relative decay of
the real and reciprocal space sums.
The last term is the so called self correction term. When evaluating
the potential at a charge location, no contribution from this charge
itself should be included, and this term is added for this purpose. 
The Ewald sums for the energy and electrostatic force are easily
obtained from the expression for the potential, see e.g. Deserno and
Holm \cite{Deserno1998}. 

Ewald sums have also been derived for the doubly and singly periodic
cases. We shall denote the situation when periodicity applies in two
dimensions and the third dimension is free as {\em planar periodicity}
or {\em 2P}. This situation is sometimes referred to as {\em
  slab/slab-like geometry} or a {\em 2d-periodic} system in
the literature. 
The Ewald sum for this case was derived e.g. by Grzybowski et
al. \cite{Grzybowski2000}.
They used an integral representation of the gamma function combined with
Poisson's summation formula, as well as a convergence factor
approach introduced in a classical derivation of the Ewald $3P$ sum by
de Leeuw et al. \cite{Leeuw1980}.
The Ewald $2P$ sum can however, as Grzybowski et al. point out, be
obtained also for example from much earlier work by Bertaut \cite{Bertaut1952}. 
Other early contributions are those by Parry
\cite{Parry1975,Parry1976} and by Leeuw and Perram \cite{Leeuw1979}.
The Ewald sum for the singly periodic, or $1P$,  case (sometimes referred to as
the $1d-periodic$ case) was derived by Porto \cite{Porto2000},
however leaving an integral expression for which no closed form was
given. A closed form can however be obtained following \cite{Fripiat2010}. 

In \cite{Lindbo2012}, we gave an alternative derivation for the $2P$
Ewald sum. This derivation was based on using Fourier series in the
periodic directions, and a Fourier integral in the free
direction. Evaluating the integral for all non-zero discrete wave
numbers, the previously derived Ewald $2P$ sum is obtained. 
The integral form of the expression was however used as the basis for
a spectrally accurate fast FFT based method for the evaluation of the 
contribution from the reciprocal space. This $2P$ Spectral Ewald method has a close
correspondance to the  Spectral Ewald method developed previously for the $3P$ case
\cite{Lindbo2011}. 

In this paper, the derivations of the Ewald $2P$ and $1P$
sums are presented in a unified framework. We do not believe that such a derivation
of the $1P$ sum has been presented before.  This
derivation illuminates the structure of the problem as well as
gives a natural starting point for the design of a fast method. 
The construction of such a method will however not be discussed here,
although a fast method will be needed in any practical implementation using an
Ewald approach, to avoid the $O(N^2)$ complexity of directly evaluating
the Ewald sums. 

The outline of the paper is as follows. 
We start by introducing the underlying idea of Ewald decomposition, whereafter
we derive the real space sum and the $k$-space sum for the $3P$ case. 
We then consider the Fourier treatment for the $2P$ case, followed by
the $1P$ case.  First we
derive the Fourier representation of the full solution (i.e without any
Ewald decomposition applied), and thereafter
the Ewald $k$-space sum.
The pure Fourier representation is used in the derivation of the
Ewald Fourier sum, to determine the terms to be added to the basic
sum in which the discrete zero mode in the periodic direction(s) is
excluded. 
The resulting Ewald summation formulas are summarized in section 
\ref{sec:ewald_formulas}. 



\section{Ewald decomposition}
\label{sect:Ewald}

The Ewald summation formula can be derived in several ways. Here, we
will utilize the fact that the electrostatic potential can 
be found as the solution to the Poisson equation
\begin{align}
-\Delta \phi =  4 \pi f^{DP}(\x), \quad f^{DP}(\x)=\sum_{\p \in P_D} \sum_n q_n \delta (\x-\x_n + \p),  \quad \x \in \R^3.
\label{eqn:laplace}
\end{align}
The sum over $\p$ is a replication of the charges in the
periodic directions, and $D=1,2,3$ indicates the number of periodic
directions with $P_D$ defined in (\ref{eqn:perdef}). 


The Ewald summation formula can be derived by 
introducing a charge screening function, $\gamma(\xi,\x)$. 
With this, we decompose $f^{DP}$ into two parts:
\begin{align*}
  f^{DP}(\x) = \underbrace{f^{DP}(\x) - (f^{DP}* \gamma)(\x)}_{:=
    f^{DP,R}(\xi,\x)} + \underbrace{(f^{DP}* \gamma)(\x)}_{:=
    f^{DP,F}(\xi,\x)}.
\end{align*}
The Poisson equation can be solved for each of the two parts of the 
right hand side to find $\phi^{PD,R}$ and $\phi^{PD,F}$, that can then 
be added.  
The screening function for which the classical Ewald decomposition in 
(\ref{eqn:Ewald3P}) is obtained is a Gaussian  $\gamma(\xi,\x)$, with 
the Fourier transform $\widehat{\gamma}(\xi,\k)$,  
\begin{align}
  \gamma(\xi,\x) = \xi^3 \pi^{-3/2} e^{-\xi^2 |\x|^2}, \quad
  \widehat{\gamma}(\xi,\k) = e^{-|\k|^2/4\xi^2}. 
\label{eq:gamma}
\end{align}
The function  $f^{DP,F}(\xi,\x)$ is smooth, and a Fourier representation
of the solution $\phi^{PD,F}$ will hence converge rapidly. 

\section{ The Ewald real space sum. }
\label{sec:Ewald_real}  

The Green's function or fundamental solution is the solution to 
$- \Delta G = 4 \pi \delta(\x)$, which yields $G(\x)=1/|\x|$.
Now, we want to find the solution to
\begin{align*}
-\Delta \phi^{DP,R}  & =  4 \pi  f^{DP,R}(\xi,\x)= 4\pi \left( f^{DP}(\x) - (f^{DP}*
\gamma)(\x) \right) \\
& = 4 \pi \sum_{n=1}^N \sum_{\p \in P_D} q_n \left( \delta(\x-\x_n)-\gamma(\x-\x_n)  \right).
\end{align*}   
To build this solution, we consider
\[
-\Delta u_n =  4 \pi \gamma(\x-\x_n) 
\]
such that $u_n(\x,\xi)$ is given by the convolution integral
\[
u_n(\x,\xi)=\int_{\reals^3} \frac{\gamma(\y-\x_n)}{|\x - \y|} \, d\y = 
\xi^3 \pi^{-3/2} \int_{\reals^3} \frac{e^{-\xi^2|\y-\x_n|^2}}{|\x - \y|} \, d\y.
\]
This convolution integral can be evaluated by introducing a spherical
coordinate system with the polar axis aligned with $\x-\x_n$, see appendix
\ref{app:conv_int_real_space}.
 The result is 
\[
u_n(\x,\xi)=\frac{\erf(\xi | \x - \x_n|) } {| \x- \x_n |}, 
\]
with $\erf()$ the error function. 
By superposition, we get
\begin{align}
\phi^{DP,R}(\x,\xi)& =\sum_n \sum_{\p \in P_D} q_n 
\left(  \frac{1} {| \x- \x_n + \p |} - \frac{\erf(\xi |\x - \x_n + \p
    | ) } {| \x- \x_n + \p |} \right), 
\notag \\
&= \sum_n \sum_{\p \in P_D} q_n \frac{\erfc(\xi | \x - \x_n + \p | ) } {| \x- \x_n + \p |}, 
\label{eqn:phi_PD_R}
\end{align}
where $\erfc(z)=1-\erf(z)$ is the complimentary error function, and $P_D$ is defined in (\ref{eqn:perdef}).
Hence, the terms in the real space sum are the same in the $1P$, $2P$
and $3P$ cases, only the periodic replication of charges is
different as reflected in the definition of  $P_D$.

To evaluate the sums at the location of a charge, $\x_m$, the
contribution of this charge must be subtracted, there is not supposed
to be any self contribution.  Simply excluding the term for $n=m$,
$\p={\bf 0}$ will not remove the full contribution, a part of it has
been included in the Fourier sum due to the decomposition. 
We compute
\[
\lim_{|\x| \rightarrow 0} \left( \frac{\erfc(\xi | \x| ) } {| \x|}-  \frac{1} {| \x|}  \right) =
\lim_{|\x| \rightarrow 0} \frac{-\erf(\xi | \x| ) } {| \x|} = -\frac{2 \xi}{\sqrt{\pi}}.
\]
This will be added to the sum multiplied by the charge strength at $\x_m$, as
can be seen for the triply periodic (3P) case in (\ref{eqn:Ewald3P}).

\section{ The k-space sum for triply periodic domains. }
\label{sec:Ewald_k_3P}

Consider $f^{3P}(\x)$ as defined in (\ref{eqn:laplace}) together with (\ref{eqn:perdef}).
Using the Poisson summation formula (Eqn. (\ref{eqn:poisson_summation}) in
appendix \ref{app:poisson_sum}),
we have that
\[
f^{3P}(\x)=
\sum_{\k} \hat{f}(\k) e^{i \k
  \cdot \x},  \quad 
\mbox{ where } \quad 
\hat{f}_{\k}= \frac{1}{V} 
\sum_n q_n e^{-i \k \cdot \x_n}, 
\]
The $\hat{f}_{\k}$ is the Fourier transform of the
term inside the periodic sum of $f^{3P}(\x)$ in (\ref{eqn:laplace}) ,
scaled with $1/V$.
We have $V=L_1L_2L_3$, 
$\k\in \{ 2\pi ( \frac{n_1}{L_1}, \frac{n_2}{L_2}, \frac{n_3}{L_3} ):
(n_1,n_2,n_3) \in Z^3 \}$ and $\hat{f}_0=0$ due to charge neutrality.

Now, expand the solution in a Fourier series as well and insert into
equation (\ref{eqn:laplace}). 
For $\k \ne 0$, we have 
\[
\hat{\phi}^{3P}_{\k} = 
\frac{4\pi}{k^2} \hat{f}_{\k}, \quad k=|\k|.
\]
For the triply periodic problem $\phi^{3P}$ will be determined only up to a constant,
which will be chosen such that 
$\int_{\Omega} \phi^{3P}(\x) d \x =0$, i.e. $\hat{\phi}^{3P}_{0}=0$.
Hence, we have 
\begin{align*}
\phi^{3P}(\x) & = \sum_{\k \ne 0} \hat{\phi}^{3P}_{\k} e^{i \k \cdot \x} = 
\sum_{\k \ne 0}  \left( \frac{4 \pi}{V} \frac{1}{k^2} \sum_n q_n e^{-i \k \cdot \x_n} \right) e^{i \k \cdot \x} = 
\frac{4 \pi}{V} \sum_{\k \ne 0}   \frac{1}{k^2} \sum_n q_n   e^{-i\k \cdot (\x_n - \x)} \\
& = \frac{4 \pi}{V}  \sum_{\k \ne 0}   \frac{1}{k^2} \sum_n q_n   e^{-i\k \cdot (\x - \x_n)}.
\end{align*}
This is the full solution to the problem, but it converges 
slowly with $k$. 

To derive the Ewald $k$ space sum, we repeat the procedure with
$f^{3P}(\x)$ replaced by $f^{3P,F}(\xi,\x)$.  
The Fourier transform of $\gamma$ is given in (\ref{eq:gamma}),
and using the fact that a convolution in real space is equal to a product
in Fourier space the Fourier coefficients of  $f^{3P,F}(\xi,\x)$ are
\[
\hat{f}^{3P,F}_{\k}= 
\frac{1}{V}
\sum_n q_n e^{-k^2/4\xi^2} e^{-i \k \cdot \x_n}, 
\]
where $k=|\k|$, 
such that 
\begin{equation}
\phi^{3P,F}(\xi,\x)= \frac{4\pi}{V} \sum_{\k \ne 0}   \frac{1}{k^2} \sum_n q_n e^{-k^2/4\xi^2}  e^{-i\k \cdot (\x - \x_n)}.
\label{eqn:3P_kspace}
\end{equation}
This is the Ewald k-space sum, as given in (\ref{eqn:Ewald3P}). 

\section{ The full k-space sum for doubly periodic domains. }
\label{sec:k_2P}

We will now again consider the Poisson equation (\ref{eqn:laplace}), but this time with periodic boundary 
conditions  in $x$ and $y$ but not in $z$, i.e. the $2P$ case with $f^{2P}(\x)$ as the right hand side. 
We will continue to use $\x$, $\x_n$, $\k$ and $k=|\k|$ as in the
previous section, but with $\k=(k_1,k_2,\kappa_3)$ to emphasize the
non-periodicity of the $z$-direction. 
We also introduce
\begin{align*}
 \r=(x,y), \quad \r_n=(x_n,y_n), \quad \kbar=(k_1,k_2), \quad \bar{k}=|\kbar|.
\label{eqn:notation2p}
\end{align*}

Expand $\phi^{2P}(\x)$ in a Fourier series in the periodic $x$ and
$y$ directions, 
\begin{equation}
\phi^{2P}(\x)=\sum_{\kbar} \hat{\phi}_{\kbar}(z) e^{i\kbar \cdot \r}. 
\label{eqn:phi2P_exp_kbar}
\end{equation}
The  $\kbar$-vectors form the discrete set 
$\{ 2\pi ( \frac{n_1}{L_1}, \frac{n_2}{L_2}): (n_1,n_2) \in Z^2 \}$, 
where $L_1$ and $L_2$ are the periodic lengths in the $x$ and $y$
directions, respectively. 
Similarly, we expand also $f^{2P}(\x)$ in (\ref{eqn:laplace}) in a Fourier series
in $x$ and $y$. The Poisson summation
formula in appendix \ref{app:poisson_sum} yields
\[
\hat{f}_{\kbar}(z)=\frac{1}{L_1 L_2} \sum_{n=1}^N q_n \delta(z - z_n)
e^{-i\kbar \cdot \r_n}. 
\]
We now insert the expansions of $\phi^{2P}(\x)$ and $f^{2P}(\x)$ into
the equation  (\ref{eqn:laplace}). Using
orthogonality of the complex exponentials, for each wave vector $\kbar$
we obtain
\[
(-\partial_z^2+\bar{k}^2)  \hat{\phi}_{\kbar}(z)= 4 \pi
\hat{f}_{\kbar}(z),  
\]
where $\bar{k}^2=|\kbar|^2=k_1^2+k_2^2$. 

The fundamental solution to this equation, i.e the solution to 
\[
(-\partial_z^2+\bar{k}^2) G(\bar{k},z)= \delta(z), 
\]
is
\begin{equation}
G(\bar{k},z)= \left\{ \begin{array}{rl} 
\frac{1}{2\bar{k}} e^{-\bar{k}|z|}, & \bar{k} \ne    0, \\ 
- \frac{1}{2} |z|, & \bar{k} = 0, 
\end{array}
\right.
\end{equation}
see e.g. \cite{Genovese2007}. 
This yields
\begin{align}
\hat{\phi}_{\kbar}(z) & =   \frac{2 \pi}{L_1 L_2}
\sum_{n=1}^N q_n \frac{1}{\bar{k}}  e^{-\bar{k}|z-z_n|} e^{-i \kbar
  \cdot \r_n}, \quad  \kbar \ne    {\bf 0}, \\ 
\hat{\phi}_{{\bf 0}}(z) & = - \frac{2 \pi}{L_1 L_2}
 \sum_{n=1}^N q_n  \left| z-z_n \right|, 
\end{align}
and hence in total we get
\begin{equation}
\phi^{2P}(\x) = \frac{2\pi}{L_1 L_2} 
\sum_{\kbar \ne 0} \sum_{n=1}^N q_n \frac{1}{\bar{k}}
e^{-\bar{k}|z-z_n|} e^{-i \kbar \cdot (\r -\r_n)}
-\frac{2\pi}{L_1 L_2} \sum_{n=1}^N q_n  \left| z-z_n \right|. 
\label{eqn:phi2P_pure_fourier}
\end{equation}
We have shifted the sign in the exponent of the complex exponential,
which does not change the sum. 
The terms in the first sum decay exponentially as $z \rightarrow \pm
\infty$ , and using charge neutrality, from the second sum we obtain 
\begin{equation}
\lim_{z  \rightarrow \pm \infty} \phi^{2P}(\x) = \pm  \frac{2\pi}{L_1 L_2}
\sum_n q_n z_n, 
\label{eqn:lim_dipole_mom}
\end{equation}
where this sum is the dipole moment in the $z$-direction. 

Now, let us consider an alternative derivation.  The Fourier
coefficients $\hat{\phi}_{\kbar}(z)$ in (\ref{eqn:phi2P_exp_kbar}) can be
represented in terms of a Fourier transform in the non-periodic
coordinate $z$, i.e. 
\begin{equation}
\hat{\phi}_{\kbar}(z) = \frac{1}{2\pi} \int_{\reals}
\tilde{\phi}_{\k} e^{i \kappa_3  z} d \kappa_3,
\label{eqn:phi_hat_kbar}
\end{equation}
where we use the notation $\k=(k_1,k_2,\kappa_3)$ to index
$\tilde{\phi}_{\k}$, although it is defined only for discrete values of
$k_1$, $k_2$ but for the continuous spectrum in $\kappa_3$. 
With this, the representation for $\phi^{2P}(\x)$ in (\ref{eqn:phi2P_exp_kbar}) can be written 
\begin{equation}
\phi^{2P}(\x)= \frac{1}{2\pi} \sum_{\kbar} \left[ \int_{\reals}
\tilde{\phi}_{\k} e^{i \kappa_3 z} d \kappa_3 \right] e^{i \kbar \cdot
\r} =
\frac{1}{2\pi} \sum_{\kbar}  \int_{\reals} \tilde{\phi}_{\k} e^{i \k \cdot \x} d \kappa_3.
\label{eqn:phi2P_mixed_exp}
\end{equation}
We represent also $f^{2P}(\x)$ this way, 
\[
f^{2P}(\x)= 
\frac{1}{2\pi} \sum_{\kbar}  \int_{\reals^2} \tilde{f}_{\k} e^{i \k \cdot \x} d \kappab, 
\]
where $\tilde{f}_{\k}$ has the same relation to $\hat{f}_{\kbar}$ as is
given in (\ref{eqn:phi_hat_kbar}) for $\phi$, or inversely
\[
\tilde{f}_{\k}= \int_{\reals} \hat{f}_{\kbar}(z) e^{-i \kappa_3 z} 
\, d z= \frac{1}{L_1 L_2} \sum_{n=1}^N q_n  e^{-i\k \cdot \x_n}. 
\]

We will now exclude the $\kbar=0$ term and consider
\begin{equation}
\bar{\phi}^{2P}(\x)= 
\frac{1}{2\pi} \sum_{\kbar \ne 0}  \int_{\reals} \tilde{\phi}_{\k} e^{i \k \cdot \x} d \kappa_3.
\label{eqn:phi2P_mixed_exp_kne0}
\end{equation}
For $\k \ne 0$, from equation (\ref{eqn:laplace}),  we have 
\[
\tilde{\phi}_{\k} = \frac{4\pi}{k^2} \tilde{f}_{\k}, 
\]
and inserting into (\ref{eqn:phi2P_mixed_exp_kne0}), we get
\begin{align}
\bar{\phi}^{2P}(\x) & = \frac{2}{L_1 L_2} \sum_{\kbar \ne 0}  \int_{\reals}
\frac{1}{k^2}  \tilde{f}_{\k} \, e^{i \k \cdot \x} \, d \kappa_3
= \frac{2}{L_1 L_2} \sum_{\kbar \ne 0}  \int_{\reals}  \frac{1}{k^2} 
\sum_{n=1}^N  q_n e^{-i \k \cdot (\x_n-\x)}  \, d \kappa_3,
\notag \\
& = \frac{2}{L_1 L_2} \sum_{\kbar \ne 0} \sum_{n=1}^N q_n  \int_{\reals}  \frac{1}{k^2} 
e^{-i \k \cdot (\x-\x_n)}  \, d \kappa_3
\label{eqn:phibar2P_pureF}
\end{align}
where 
$\k=(k_1,k_2,\kappa_3)$ and $k=|\k|$.  
The Fourier vectors $\kbar$ form the discrete set 
$\{ 2\pi ( \frac{n_1}{L_1}, \frac{n_2}{L_2} ): (n_1,n_2) \in Z^2 \}$. 
Hence, the expression is the same as in the triply periodic case, with
the Fourier sum replaced by a Fourier integral in the non-periodic
$z$  direction. 

Expanding the terms containing $\k$ and $k$, we have
\begin{equation}
\bar{\phi}^{2P}(\x) = 
\frac{2}{L_1 L_2} \sum_{\kbar \ne 0}  \sum_{n=1}^N q_n 
e^{-i \kbar \cdot (\r - \r_n)}  
\int_{\reals}  \frac{1}{\bar{k}^2+ \kappa_3^2} 
e^{-i\kappa_3  (z - z_n)}  \, d \kappa_3, 
\label{eqn:phibar2P_pureF_exp}
\end{equation}
where $\x=(x,y,z)$, $\r=(x,y)$, $\bar{k}^2=|\kbar|^2=k_1^2+k_2^2$.
We can evaluate 
\begin{equation}
\bar{I}(\bar{k},z)=  \int_{-\infty}^{\infty}  \frac{1}{\bar{k}^2+ \kappa_3^2} 
e^{-i \kappa_3 z}  \, d \kappa_3  = \pi \frac{1}{\bar{k}} e^{-\bar{k} |z|}.
\label{eqn:PureFourierInt2P}
\end{equation}
Hence, we get
\begin{align*}
\bar{\phi}^{2P}(\x) = 
\frac{2 \pi}{L_1 L_2} \sum_{\kbar \ne 0}  \sum_{n=1}^N q_n 
 \frac{1}{\bar{k}} e^{-\bar{k} |z
    - z_n|} e^{-i \kbar \cdot  (\r - \r_n)}.
\end{align*}
This is the same expression as was obtained in
(\ref{eqn:phi2P_pure_fourier}). 
This sum has a slow convergence in $\bar{k}$ for $z$ close to any
$z_n$. 

\section{ The Ewald k-space sum for doubly periodic domains. }
\label{sec:Ewald_k_2P}

For the derivation of the Ewald k-space sum, it is most convenient to follow the second path
of derivation from above. 
This modifies the expression for $\bar{\phi}^{2P}$ in (\ref{eqn:phibar2P_pureF})
with a Gaussian term, and we have 
\begin{align*}
\bar{\phi}^{2P,F}(\x,\xi) = 
\frac{2}{L_1 L_2} \sum_{\kbar \ne 0} \sum_{n=1}^N q_n  \int_{\reals}  \frac{1}{k^2} 
e^{-k^2/4\xi^2}  e^{-i \k \cdot (\x-\x_n)}  \, d \kappa_3
\end{align*}
where we have introduced the superscript $F$ for this term. 
Compare to the introduction of the Gaussian factor in 
(\ref{eqn:3P_kspace}). 

This sum excludes the term for $\kbar=0$, similarly to the definition
for $\bar{\phi}^{2P}$ for the pure Fourier sum, and such a term must
be added. 
In total, we write
\begin{equation}
\phi^{2P}(\x)=\phi^{2P,R}(\x,\xi)+\bar{\phi}^{2P,F}(\x,\xi)+\bar{\phi}^{F,\kbar=0}(\x,\xi), 
\label{eqn:phi2P_tot_def}
\end{equation}
where $\phi^{2P,R}(\x,\xi)$ is the real space sum.
Once the potential is to be evaluated at the location of a charge, the
contribution from that charge should not be included. We will make
this correction at the end, in section \ref{sec:ewald_formulas}. 

Expanding the $\k$ vector in the expression above for $\bar{\phi}^{2P,F}(\x,\xi)$,  we find
\begin{align}
\bar{\phi}^{2P,F}(\x) = 
\frac{2}{L_1 L_2} \sum_{\kbar \ne 0}  \sum_{n=1}^N q_n 
e^{-i \kbar \cdot (\r - \r_n)}  
\int_{\reals}  \frac{e^{-(\bar{k}^2+\kappa_3^2)/4\xi^2} }{\bar{k}^2+ \kappa_3^2} 
e^{-i\kappa_3  (z - z_n)}  \, d \kappa_3.
\label{eqn:phibar2P_F_exp}
\end{align}
From \cite{Gradshteyn2007} (3.954 (2), p. 504), 
\begin{align*}
\int_{\R}  \frac{e^{-(\bar{k}^2+\kappa^2)/4\xi^2}}{\bar{k}^2+\kappa^2} e^{-i\kappa z}  \, d \kappa= 
\frac{\pi}{2} \frac{1}{\bar{k}}\bigg[ e^{\bar{k}z}
  \erfc\left(\frac{\bar{k}}{2\xi} + \xi z\right)
  + e^{-\bar{k}z} \erfc\left(\frac{\bar{k}}{2\xi} - \xi z\right)
  \bigg ], 
\end{align*}
Introducing
\begin{equation}
g(\bar{k},z,\xi)=e^{\bar{k}z}
  \erfc\left(\frac{\bar{k}}{2\xi} + \xi z\right)
  + e^{-\bar{k}z} \erfc\left(\frac{\bar{k}}{2\xi} - \xi z\right), 
\label{eq:g_def_w_erfc}
\end{equation}
we get
\begin{align}
\bar{\phi}^{2P,F}(\x) = 
\frac{\pi}{L_1 L_2} 
\sum_{n=1}^N q_n 
\sum_{\kbar \ne 0}  
e^{-i \kbar \cdot (\r - \r_n)}\frac{1}{\bar{k}}  
g(\bar{k},z-z_n,\xi).
\label{eqn:phibar2P_F_final}
\end{align}

This sum excludes the term for $\kbar=0$. The $\kbar=0$ term for the pure Fourier expression
is given by the second term in (\ref{eqn:phi2P_pure_fourier}). 
Due to the Ewald decomposition, some of the $\kbar=0$ mode will
however be included into the real space term,  
and $\bar{\phi}^{F,\kbar=0}(\x,\xi)$ equals the $\kbar=0$ term for the
pure Fourier expression with the contribution from the real space sum
removed. This real space contribution is most accessible as the
difference between the two Fourier expressions, in the limit of zero
wavenumber. 
Hence, we define
\begin{align}
\bar{\phi}^{F,\kbar=0}(\x,\xi)   = &
-\frac{2\pi}{L_1 L_2} \sum_{n=1}^N q_n  \left| z-z_n \right| 
\notag \\
& - \frac{\pi}{L_1 L_2} \sum_{n=1}^N q_n \lim_{\kbar\rightarrow 0} 
\left[ e^{-i\kbar \cdot (r - r_n)}  \frac{1}{\bar{k}} 
\left( g(\bar{k},z-z_n,\xi)  - 2 e^{-\bar{k}|z|} \right) \right].
\label{eqn:phiF_k0_def2P}
\end{align}
We can compute the limit (see appendix \ref{app:lim_k0_2P} for details)
\[
\lim_{\bar{k}\rightarrow 0}  \left[  \frac{1}{\bar{k}} 
\left( g(\bar{k},z-z_n,\xi)  - 2 e^{-\bar{k}|z|} \right) \right]
=-2 \left(  
\frac{1}{\xi \sqrt{\pi}} e^{-(\xi z)^2}-|z| +z \, \erf(\xi z) 
\right)
\]
and in total we get
\begin{align}
 \phi^{F,\kbar=0}(\x,\xi)=-\frac{2 \sqrt{\pi}}{L_1 L_2} \sum_{n=1}^N q_n
\left(  \frac{1}{\xi} e^{-(\xi (z-z_n))^2} +\sqrt{\pi} (z-z_n) \erf(\xi (z-z_n)) \right).
\label{eqn:phiF_k0_2P}
\end{align}
With this and the result for $\bar{\phi}^{2P,F}(\x)$ in (\ref{eqn:phibar2P_F_final}), we have defined the 
k-space terms in (\ref{eqn:phi2P_tot_def}). The full Ewald sum will be
stated in section \ref{sec:ewald_formulas}.

\section{ The full k-space sum for singly periodic domains. }
\label{sec:k_1P}

We will now again consider the Poisson equation (\ref{eqn:laplace}), but this time with periodic boundary 
conditions only in $z$ and "free space" in $x$ and $y$, i.e. the 1P
case, with $f^{1P}(\x)$ as the right hand side. 
We start by expanding $\phi^{1P}(\x)$ in a Fourier series in the periodic $z$-direction, 
\begin{equation}
\phi^{1P}(\x)=\sum_{k_3} \hat{\phi}_{k_3}(\r) e^{ik_3z}. 
\label{eqn:phi_exp_k3}
\end{equation}
The Fourier modes $k_3$ form the discrete set 
$\{ 2\pi n/L_3:  n \in Z \}$, where $L_3$ is the periodic length in the $z$ direction.   
We now expand also $f^{1P}(\x)$ in a Fourier series, and use the Poisson summation
formula in appendix \ref{app:poisson_sum} to obtain
\[
\hat{f}_{k_3}(\r)=\frac{1}{L_3} \sum_{n=1}^N q_n \delta(\r - \r_n) e^{-ik_3z_n}. 
\]
By inserting the expansions for $\phi^{1P}(\x)$ and $f^{1P}(\x)$
into  (\ref{eqn:laplace}), and using
orthogonality of the complex exponentials, for each wave number $k_3$
we obtain
\[
(-\Delta_{2D}+k_3^2)  \hat{\phi}_{k_3}(\r)= 4 \pi \hat{f}_{k_3}(\r), 
\]
where $\Delta_{2D}$ denotes the Laplacian in the $xy$-plane. This PDE
is the screened Poisson equation. The fundamental solution to 
this equation, i.e the solution to 
\[
(-\Delta_{2D}+k_3^2)  G(\r)= \delta(\r), 
\]
is
\begin{equation}
G(\r)= \left\{ \begin{array}{ll} 
\frac{1}{2\pi} K_0(|k_3| \rho), & k_3 \ne    0, \\ 
- \frac{1}{2\pi} \log(\rho), & k_3 = 0, 
\end{array}
\right.
\end{equation}
where $\rho=|\r|=\sqrt{x^2+y^2}$, and where $K_0$ is the modified Bessel function of
the second kind (available as \texttt{besselk} in Matlab). 

This yields
\begin{align}
\hat{\phi}_{k_3}(\r) & = \frac{2}{L_3} \sum_{n=1}^N q_n K_0(|k_3|
\rho_n) e^{-ik_3z_n}, \quad  k_3 \ne    0, \\ 
\hat{\phi}_{0}(\r) & = -\frac{2}{L_3} \sum_{n=1}^N q_n  \log(\rho_n), 
\end{align}
and hence in total we get
\begin{equation}
\phi^{1P}(\x) = \frac{2}{L_3} 
\sum_{k_3 \ne 0} \sum_{n=1}^N q_n e^{-ik_3 (z - z_n)} K_0(|k_3|
\rho_n)  - \frac{1}{L_3} \sum_{n=1}^N q_n  \log(\rho_n^2)
\label{eqn:phi1P_pure_fourier}
\end{equation}
where $\rho_n=|\r-\r_n|=\sqrt{(x-x_n)^2+(y-y_n)^2}$.
For small arguments $z>0$, it holds that
(\cite{Abramowitz1970}, p 375, 9.6.13), 
\[
K_0(z) \approx -\log(z/2)-\gamma,
\]
where $\gamma$ is the Euler–Mascheroni constant ($0.5772...$). 
Hence, both terms in (\ref{eqn:phi1P_pure_fourier}) have a logarithmic singularity at $\rho_n=0$.
However, as is shown in \cite{Grzybowski2002}, the total expression is
indeed finite as long as $z \ne z_n$. 
For large arguments, $z>0$, we have the expansion
(\cite{Abramowitz1970}, p 378, 9.7.2), 
\[
K_0(z) \approx \sqrt{\frac{2}{\pi z}} e^{-z} \left(1-\frac{1}{8z}+\frac{9}{2
  (8z)^2} + \cdots \right)
\]
Hence, there is an exponential decay of the terms in the first sum,
but for small values of $\rho_n$ this decay will be very slow in
$|k_3|$.
From the second sum, it looks as if the solution has a logarithmic
growth in $\rho_n$. However, due to charge neutrality, it actually
decays as $r=\sqrt{x^2+y^2} \rightarrow \infty$. 
For details, see  \ref{app:logsum}. 

Now, let us similarly to the 2P case consider an alternative
derivation.  We will denote $\k=(\kappa_1,\kappa_2,k_3)$, again using
$\kappa$ in the free directions, and also introduce
$\kappab=(\kappa_1,\kappa_2)$. The Fourier
coefficients $\hat{\phi}_{k_3}(\r)$ in (\ref{eqn:phi_exp_k3}) can now be
represented in terms of a Fourier transform in the non-periodic
coordinates $x$ and $y$, i.e. 
\begin{equation}
\hat{\phi}_{k_3}(\r) = \frac{1}{(2\pi)^2} \int_{\reals^2}
\tilde{\phi}_{\k} e^{i \kappab \cdot \r} d \kappab, 
\label{eqn:phi_hat_k3}
\end{equation}
where we use $\k=(\kappa_1,\kappa_2,k_3)$ to index
$\tilde{\phi}_{\k}$, although it is defined only for discrete values of
$k_3$ but for the continuous spectrum in $\kappa_1,\kappa_2$. 
Inserting into the definition of $\phi^{1P}(\x)$ in
(\ref{eqn:phi_exp_k3}), we get
\begin{equation}
\phi^{1P}(\x)= \frac{1}{(2\pi)^2} \sum_{k_3} \left[ \int_{\reals^2}
\tilde{\phi}_{\k} e^{i \kappab \cdot \r} d \kappab \right] e^{ik_3z} =
\frac{1}{(2\pi)^2} \sum_{k_3}  \int_{\reals^2} \tilde{\phi}_{\k} e^{i \k \cdot \x} d \kappab. 
\label{eqn:phi1P_mixed_exp}
\end{equation}
Similarly, 
\[
f^{1P}(\x)= 
\frac{1}{(2\pi)^2} \sum_{k_3}  \int_{\reals^2} \tilde{f}_{\k} e^{i \k \cdot \x} d \kappab 
\]
where $\tilde{f}_{\k}$ has the same relation to $\hat{f}_{k3}$ as is
given in (\ref{eqn:phi_hat_k3}) for $\phi^{1P}$, or inversely
\[
\tilde{f}_{\k}= \int_{\reals^2} \hat{f}_{k3}(\r) e^{-i \kappab \cdot
  \r} \, d \r= \frac{1}{L_3} \sum_{n=1}^N q_n  e^{-i\k \cdot \x_n}. 
\]

We will now exclude the $k_3=0$ term and consider
\begin{equation}
\bar{\phi}^{1P}(\x)= 
\frac{1}{(2\pi)^2} \sum_{k_3 \ne 0}  \int_{\reals^2} \tilde{\phi}_{\k}
e^{i \k \cdot \x} d \kappab. 
\label{eqn:phi1P_mixed_exp_kne0}
\end{equation}
For $\k \ne 0$, from equation (\ref{eqn:laplace}),  we have the same
relation as previously, 
$\tilde{\phi}_{\k} = \frac{4\pi}{k^2} \tilde{f}_{\k}$, 
and inserting into (\ref{eqn:phi1P_mixed_exp_kne0}), we get
\begin{align}
\bar{\phi}^{1P}(\x) & = \frac{1}{\pi} \sum_{k_3 \ne 0}  \int_{\reals^2} \int_{\R}
\frac{1}{k^2}  \tilde{f}_{\k} \, e^{i \k \cdot \x} \, d \kappab
= \frac{1}{\pi L_3} \sum_{k_3 \ne 0}  \int_{\reals^2}  \frac{1}{k^2} 
\sum_{n=1}^N  q_n e^{-i \k \cdot (\x_n-\x)}  \, d \kappab,
\notag \\
& = \frac{1}{\pi L_3} \sum_{k_3 \ne 0} \sum_{n=1}^N q_n  \int_{\reals^2}  \frac{1}{k^2} 
e^{-i \k \cdot (\x-\x_n)}  \, d \kappab,
\label{eqn:phibar1P_pureF}
\end{align}
where 
$\k=(\kappa_1,\kappa_2,k_3)$ and $k=|\k|$.  
The Fourier modes $k_3$ form the discrete set 
$\{ 2\pi n/L_3:  n \in Z \}$, where $L_3$ is the periodic length in the $z$ direction.   
Hence, the expression is the same (modulo a constant) as compared to the triply and double periodic
cases, but in each case we have Fourier sums in the periodic directions 
and Fourier integrals in the non-periodic ones. 

Expanding the terms containing $\k$ and $k$, we have
\begin{equation}
\bar{\phi}^{1P}(\x) = 
\frac{1}{\pi L_3} 
\sum_{k_3 \ne 0} \sum_{n=1}^N q_n e^{-ik_3 (z - z_n)}  \int_{\R^2}   \frac{1}{\kappa_1^2+\kappa_2^2+ k_3^2} 
e^{-i\kappab \cdot (\r - \r_n)}  \, d \kappa_1 d\kappa_2 \  , 
\label{eqn:phibar1P_pureF_exp}
\end{equation}
where $\x=(x,y,z)$, $\r=(x,y)$, $\kappab=(\kappa_1,\kappa_2)$.

We can evaluate 
\begin{equation}
\bar{I}(k,x,y)= \int_{-\infty}^{\infty} \int_{-\infty}^{\infty}  \frac{1}{\kappa_1^2+\kappa_2^2+ k^2} 
e^{-i(\kappa_1 x + \kappa_2 y)}  \, d \kappa_1 d\kappa_2 = 2\pi K_0(k\rho),
\label{eqn:PureFourierInt1P}
\end{equation}
where $\rho=\sqrt{x^2+y^2}$, and where $K_0$ is the modified Bessel function of
the second kind as was already introduced. See 
\ref{app:integrals} for details.

Hence, we get
\begin{align*}
\bar{\phi}^{1P}(\x) = 
\frac{2}{L_3} 
\sum_{k_3 \ne 0} \sum_{n=1}^N q_n e^{-ik_3 (z - z_n)} K_0(|k_3| \rho_n),
\end{align*}
where
$\rho_n=|\r-\r_n|=\sqrt{(x-x_n)^2+(y-y_n)^2}$.
This is the same expression, as was obtained in
(\ref{eqn:phi1P_pure_fourier}).

\section{ The Ewald k-space sum for singly periodic domains. }
\label{sec:Ewald_k_1P}
To derive the Ewald k-space sum, we will follow the second path
of derivation from above, as was done also for the $2P$ case. 
This will again introduce a Gaussian factor as compared to 
the expression for $\bar{\phi}^{1P}$ in (\ref{eqn:phibar1P_pureF}), 
\begin{align*}
\bar{\phi}^{1P,F}(\x,\xi) = \frac{1}{\pi L_3} \sum_{k_3 \ne 0} \sum_{n=1}^N q_n
\int_{\R^2}  \frac{1}{k^2} 
e^{-k^2/4\xi^2}  e^{-i \k \cdot (\x-\x_n)}  \, d \kappa_1 d\kappa_2, 
\end{align*}
where we have introduced the superscript $F$ for this term. 

This sum excludes the term for $k_3=0$, and this term will be derived
below, similarly to what was done for the $2P$ case. 
In total, we write
\[
\phi ^{1P} (\x)=\phi^{1P,R}(\x,\xi)+\bar{\phi}^{1P,F}(\x,\xi)+\bar{\phi}^{F,k_3=0}(\x,\xi), 
\]
where $\phi^{1P,R}(\x,\xi)$ is the real space sum, and was discussed in
section \ref{sec:Ewald_real}. 
The self correction term needed when evaluating the potential at the location
of a charge, as discussed below (\ref{eqn:phi_PD_R}), will be added in
the final equation in section \ref{sec:ewald_formulas}.


Considering the expression for $\bar{\phi}^{1P,F}(\x,\xi)$,  again expanding the $\k$ vector, we find
\begin{align}
\bar{\phi}^{1P,F}(\x,\xi) = 
\frac{1}{\pi L_3} 
\sum_{k_3 \ne 0} \sum_{n=1}^N q_n e^{-ik_3 (z - z_n)}
e^{-k_3^2/4\xi^2} \int_{\R^2}   \frac{e^{-(\kappa_1^2+\kappa_2^2)/4\xi^2} }{\kappa_1^2+\kappa_2^2+ k_3^2} 
e^{-i\kappab \cdot (\r - \r_n)}  \, d \kappa_1 d\kappa_2.
\label{eqn:phibar_F_exp}
\end{align}
Let us define
\begin{align*}
I(k_3,x,y,\xi)=e^{-k_3^2/4\xi^2} \int_{\R^2}   \frac{1}{\kappa_1^2+\kappa_2^2+ k_3^2} 
e^{-(\kappa_1^2+\kappa_2^2)/4\xi^2} e^{-i\kappab \cdot \r}  \, d\kappa_1 d\kappa_2.
\end{align*}
From the derivation in \ref{app:integrals}, we have
\begin{align*}
I(k_3,x,y,\xi)=\pi K_0(k_3^2/4\xi^2,\rho^2 \xi^2),
\end{align*}
where $\rho=\sqrt{x^2+y^2}$. 
The function $K_0(u,v)$ is an incomplete modified Bessel function of
the second kind. 
It is defined as
\begin{equation}
K_0(u,v)=\int_1^{\infty} \frac{dt}{t} e^{-ut-v/t}.
\label{eqn:K0_two_args}
\end{equation}
Note that this is not the same function as the $K_0$ of one argument introduced
above. 

To summarize, we have 
\begin{align}
\bar{\phi}^{1P,F}(\x,\xi) & = 
\frac{1}{\pi L_3} 
\sum_{k_3 \ne 0} \sum_{n=1}^N q_n e^{-ik_3 (z - z_n)}
I(k_3,x-x_n,y-y_n,\xi) \notag \\
&=
\frac{1}{ L_3} 
\sum_{k_3 \ne 0} \sum_{n=1}^N q_n e^{-ik_3 (z - z_n)}
 K_0(k_3^2/4\xi^2,\rho_n^2 \xi^2),
\label{eqn:phibar1P_F_final}
\end{align}
where $\rho_n=\sqrt{(x-x_n)^2+(y-y_n)^2}$. 

We now need to compute $\bar{\phi}^{F,k_3=0}(\x,\xi)$.
The $k_3=0$ term for the pure Fourier case is 
the sum over logarithmic terms in (\ref{eqn:phi1P_pure_fourier}).  
This sum must be corrected with the part of the  $k_3=0$ mode that has 
been included into the real space sum. Again, as in the 2P case, 
we will find this contribution as the difference of the two 
Fourier expansions in the limit of vanishing wave number. 
We define
\begin{align*}
\bar{\phi}^{F,k_3=0}(\x,\xi)  = & - \frac{1}{L_3} \sum_{n=1}^N q_n \log(\rho_n^2) \\
& - \frac{1}{L_3} \sum_{n=1}^N q_n \lim_{|k_3|\rightarrow 0} 
\left[ e^{-ik_3 (z - z_n)}  ( 2 K_0(|k_3|
  \rho_n)-K_0(k_3^2/4\xi^2,\rho_n^2 \xi^2)) \right]
\end{align*}
To compute the needed limit,  we can use the fact that, for small $u$, 
\[
K_0(u,v)=2K_0(2\sqrt{uv})-E_1(v)+O(u), 
\]
where $E_1(v)$ is the exponential integral, defined as
(\cite{Abramowitz1970}, p 228, 5.1.1), 
\begin{equation} 
E_1(v)=\int_{1}^{\infty} \frac{1}{t} e^{-vt} dt=\int_{v}^{\infty} \frac{1}{t} e^{-t} dt,
\label{eqn:expint}
\end{equation}
(and available e.g. in Matlab using \texttt{expint}).
With this, we get 
\[
 \lim_{k_3\rightarrow 0}   \left[ 2 K_0(|k_3|
   \rho)-K_0(k_3^2/4\xi^2,\rho^2 \xi^2) \right] =E_1(\rho_n^2 \xi^2),
\]
and so in total, we get
\begin{align}
\bar{\phi}^{F,k_3=0}(\x,\xi) & 
= - \frac{1}{L_3} \sum_{n=1}^N q_n \left\{ \log(\rho_n^2) +E_1(\rho_n^2
  \xi^2) \right\} \notag \\
&= \frac{1}{L_3} \sum_{n=1}^N q_n \left\{
  \gamma - \log(\rho_n^2\xi^2) -E_1(\rho_n^2 \xi^2) \right\}.
\label{eqn:phiF_k0_1P}
\end{align}
where we have used charge neutrality in the second step, see 
\ref{app:logsum}. 

The exponential integral can be expanded as
(\cite{Abramowitz1970}, p 229, 5.1.11), 
\[
E_1(x)=-\gamma-\log(x)-\sum_{p=1}^{\infty} (-1)^p \frac{x^p}{p!p}, 
\]
from which it follows that 
\begin{equation}
\lim_{\rho \rightarrow 0} \left\{
  \gamma - \log(\rho^2\xi^2) -E_1(\rho^2 \xi^2) \right\}=0. 
\label{eqn:lim_gamma_log_E1}
\end{equation}
This makes the second form in (\ref{eqn:phiF_k0_1P}) especially convenient
when evaluating at a charge location.

\section{ The Ewald summation formulas. } 
\label{sec:ewald_formulas}

The Ewald summation formula for a triply periodic, charge neutral
system has already been given in (\ref{eqn:Ewald3P}). 
For a doubly periodic system, periodic in $x$ and $y$, but not in $z$, 
$\phi^{2P,R}(\x,\xi)$ has been defined in (\ref{eqn:phi_PD_R}), 
$\bar{\phi}^{2P,F}(\x,\xi)$ in (\ref{eqn:phibar2P_F_final}) and
$\bar{\phi}^{F,\kbar=0}(\x,\xi)$ in (\ref{eqn:phiF_k0_2P}). 
The self correction term reamins the same as in the triply periodic
case, as discussed in section \ref{sec:Ewald_real}.  
Adding it all together, we get
\begin{align}
  \phi^{2P}(\x_m) =& \sum_{\p \in P_2} \sum_{n=1}^{N,'} q_n \frac{\erfc(\xi
    |\x_m - \x_n +
    \p|)}{|\x_m - \x_n + \p|} + \nonumber \\
  &+\frac{\pi}{L_1L_2}\sum_{n=1}^N \sum_{\kbar \neq 0} e^{-i \kbar \cdot
      (\r_m-\r_n)} \frac{1}{\bar{k}} g(\bar{k},z_m-z_n,\xi)
\nonumber \\
  &-\frac{2\sqrt{\pi}}{L_1 L_2} \sum_{n=1}^N q_n \left( \frac{1}{\xi} e^{-\xi^2
      (z_m-z_n)^2} + \sqrt{\pi} (z_m-z_n) \erf(\xi(z_m-z_n))
  \right) -\frac{2\xi}{\sqrt{\pi}}q_m, 
\label{eq:2P_ewald_sum}
\end{align}
where 
\[
g(\bar{k},z,\xi)= e^{\bar{k}z}
  \erfc\left(\frac{\bar{k}}{2\xi} + \xi z\right)
  + e^{-\bar{k}z} \erfc\left(\frac{\bar{k}}{2\xi} - \xi z\right).
\]
Furthermore, the ${N,'}$ indicates that the term ($n=m$, $\p={\bf 0}$)
is excluded from the real space sum, 
$P_2 = \{ (jL_1,lL_2,0)\}: (j,l) \in \Z^2 \}$, 
as defined in (\ref{eqn:perdef}) and the Fourier vectors $\kbar$ form the
discrete set $\{ 2\pi ( \frac{n_1}{L_1}, \frac{n_2}{L_2} ): (n_1,n_2)
\in Z^2 \}$.

For a system that is periodic only in the $z$ direction (the 1P
case), 
$\phi^{1P,R}(\x,\xi)$ has been defined in (\ref{eqn:phi_PD_R}),
$\bar{\phi}^{1P,F}(\x,\xi)$ in (\ref{eqn:phibar1P_F_final}) and
$\bar{\phi}^{F,k_3=0}(\x,\xi)$ in (\ref{eqn:phiF_k0_1P}). 
Also here, the self correction term remains the same.
Adding all these components, we obtain the Ewald summation formula for
a charge neutral system, periodic in the $z$-direction with a periodic
length $L_3$.  The potential evaluated at a source location $\x_m$,
$m=1,\ldots,N$ will be
\begin{align}
\phi(\x_m)= & \sum_{\p \in P_1} \sum_{n=1}^{N,'} q_n \frac{
\text{erfc}(\xi \, |\x_m-\x_n+\p |)}{|\x_m-\x_n+\p |} 
+\frac{1}{ L_3} \sum_{k_3 \ne 0} \sum_{n=1}^N q_n e^{-ik_3 (z_m - z_n)}
 K_0(k_3^2/4\xi^2,\rho_{mn}^2 \xi^2)
\notag \\
& + \frac{1}{L_3} \sum_{\shortstack{\scriptsize $n=1$ \\ \scriptsize $n \ne m$}}^N q_n \left\{
  \gamma - \log(\rho_{mn}^2\xi^2) +E_1(\rho_{mn}^2 \xi^2) \right\}
 - \frac{2 \xi}{\sqrt{\pi}}q_m,
\label{eqn:FullEwald1P}
\end{align}
where the ${N,'}$ indicates that the term ($n=m$, $\p={\bf 0}$) is excluded from the real space
sum, and $P_1=\{ (0,0,lL\}: l \in \Z^3$, as defined in (\ref{eqn:perdef}).
The Fourier modes $k_3$ form the discrete set $\{ 2\pi n/L_3: n \in Z
\}$ and $\rho_{mn}=|\r_m-\r_n|=\sqrt{(x_m-x_n)^2+(y_m-y_n)^2}$.
The function $K_0(.,.)$ is an  incomplete modified Bessel function of
the second kind, as defined in (\ref{eqn:K0_two_args}). 
The constant $\gamma$ is the Euler–Mascheroni constant ($0.5772...$)
and the function $E_1(.)$ is the exponential integral, as defined in 
(\ref{eqn:expint}). 
The term $n=m$ in the last sum can be skipped due to (\ref{eqn:lim_gamma_log_E1}).

\appendix
\section{Poisson summation formula.}
\label{app:poisson_sum}



Considering periodic sums, with $P_D$ as defined in
(\ref{eqn:perdef}), Poisson's summation formula yields the following relations
\begin{alignat}{2}
\sum_{\p \in P_3} g(\x+\p) & = \sum_{\k} \hat{g}(\k) e^{i \k
  \cdot \x}, \quad &  &
\hat{g}(\k) = \frac{1}{L_1L_2L_3} \int_{\reals^3} g(\x) e^{-i \k \cdot \x} d\x,
\notag \\
\sum_{\p \in P_2} g(\x+\p) & = \sum_{\kbar} \hat{g}(\kbar,z) e^{i \kbar
  \cdot \r},  \quad & &
\hat{g}(\kbar,z) =\frac{1}{L_1L_2} \int_{\reals^2} g(\x) e^{-i \kbar \cdot \r} d\r, 
\notag \\
\sum_{\p \in P_1} g(\x+\p) & = \sum_{k_3} \hat{g}(k_3,\r)
e^{i k_3 z}, \quad & & \hat{g}(k_3,\r) =\frac{1}{L_3}\int_{\reals} g(\x) e^{-i k_3 z} dz.
\label{eqn:poisson_summation}
\end{alignat}
Here, the $\hat{g}$ is the continuous Fourier transform, applied in
the periodic direction(s), and 
$\x=(\r,z)=(x,y,z)$, $\k=(k_1,k_2,k_3)$ and $\kbar=(k_1,k_2)$. 
Note that in the most common form of Poisson's summation formula, the
factors with $L_1$, $L_2$ and $L_3$ would appear in front of the
$k$-sums instead of in the definitions of the $\hat{g}$:s. 
This form is however more convenient for our purposes. 

\section{Convolution integral for real space sum.}
\label{app:conv_int_real_space}

We want to evaluate 
\[
u_n(\x_t,\xi) =\int_{\reals^3} \frac{\gamma(\y-\x_n)}{|\x_t - \y|} \, d\y = 
\xi^3 \pi^{-3/2} \int_{\reals^3} \frac{e^{-\xi^2|\y-\x_n|^2}}{|\x_t - \y|} \, d\y,
\]
where the Gaussian is centered at $\x_n$, and the evaluation point is
denoted $\x_t$. 

Let ${\bf R_0}=\x_t-\x_n$ and assume a coordinate system such that 
${\bf R_0}=(0,0,R_0)$, and use spherical coordinates s.t. 
$\y-\x_t=(r \sin \theta \cos \varphi, r \sin \theta \sin \varphi, r
\cos \theta)$. 
This yields $|\y-\x_n|^2=|\y-\x_t + {\bf R_0}|^2=r^2+2rR_0 \cos \theta
+R_0^2$.
Now, we will integrate over a sphere centered in $\x_t$ with radius
$B$, and at the end let the radius go to infinity. We evaluate
\begin{align*}
I(R_0,\xi,B)=\xi^3 \pi^{-3/2}\int_0^B \int_0^{\pi} \int_0^{2\pi} \frac{1}{r}
e^{-\xi^2(r^2+2rR_0 \cos \theta+R_0^2)} \, r^2 \sin \theta \, d\varphi
d\theta dr. 
\end{align*}
The integral over $\varphi$ simply yields a factor of $2\pi$, and the
integral over $\theta$ is not difficult to evaluate. 
We are left with
\begin{align*}
I(R_0,\xi,B)& =\frac{\xi}{\sqrt{\pi}} \frac{1}{R_0}
\int_0^B  \left( e^{-\xi^2(r-R_0 )^2} - e^{-\xi^2(r+R_0 )^2} \right) dr \\
& = \frac{1}{2 R_0} \left(2 \erf(\xi R_0)
  +\erf(\xi(B-R_0))-\erf(\xi(B+R_0)) \right).
\end{align*}
Using $\lim_{z \rightarrow \infty} \erf(z)=1$, we obtain
\[
u_n(\x_t,\xi) =\lim_{B \rightarrow \infty} I(| \x_t- \x_n |,\xi,B)=\frac{\erf(\xi | \x_t - \x_n|) } {| \x_t- \x_n |}. 
\]

\section{Limit for the $\kbar=0$ term in the 2P case.}
\label{app:lim_k0_2P}

We need to comute the limit
\begin{align*}
A(z,\xi) =
 \lim_{k\rightarrow 0} \frac{1}{k} \left( e^{kz}
  \erfc\left(\frac{k}{2\xi} + \xi z\right)
  + e^{-kz} \erfc\left(\frac{k}{2\xi} - \xi z\right)
 - 2 e^{-k|z|} \right). 
\end{align*}
Assuming $z>0$, we first compute
\begin{align*}
A^+(z,\xi)= \lim_{k\rightarrow 0} \frac{1}{k} \left( 
e^{kz}-e^{-kz}
-e^{kz}\erf\left(\xi z+\frac{k}{2\xi} \right)
+e^{-kz}\erf\left(\xi z-\frac{k}{2\xi} \right)
 \right). 
\end{align*}
Both nominator and denominator have zero limit, 
so we apply L'Hopitals rule. 
Differentiating the denominator simply gives $1$. 
Differentiating the first two terms in the nominator yields
\begin{align*}
z e^{kz}+ze^{-kz}
\end{align*}
and differentiating the remaining terms in the nominator yields
\begin{align*}
-ze^{kz}\erf\left(\xi z+\frac{k}{2\xi} \right)
-ze^{-kz}\erf\left(\xi z-\frac{k}{2\xi} \right)
+\frac{1}{\xi \sqrt{\pi}} \left( 
-e^{kz}e^{-\left(\xi z+\frac{k}{2\xi} \right)^2}
- e^{kz}e^{-\left(\xi z-\frac{k}{2\xi} \right)^2}
\right).
\end{align*}
With this we get 
\begin{align*}
A^+(z,\xi)= \left(  
2z -2z\erf(\xi z)  -\frac{2}{\xi \sqrt{\pi}} e^{-(\xi z)^2}
\right)
=-2  \left(  \frac{1}{\xi \sqrt{\pi}} e^{-(\xi z)^2} -z +z \erf(\xi z) \right)
\end{align*}
Performing the same calculation when $z<0$ yields
\begin{align*}
A^-(z,\xi)=
-2 \left(  
\frac{1}{\xi \sqrt{\pi}} e^{-(\xi z)^2}
+z +z \erf(\xi z) 
\right).
\end{align*}
We can write these two limits as one, and we get
\[
A(z,\xi)= -2 \left(  
\frac{1}{\xi \sqrt{\pi}} e^{-(\xi z)^2}-|z| +z \, \erf(\xi z) 
\right).
\]

\section{Evaluation of integrals in the 1P case.}
\label{app:integrals}

Consider the integral
\[
\bar{I}(k,x,y)= \int_{-\infty}^{\infty}  \int_{-\infty}^{\infty} 
\frac{1}{\kappa_1^2+ \kappa_2^2 + k^2} e^{-i( \kappa_1 x+ \kappa_2 y)}
\, d \kappa_1 \, d \kappa_2.
\]
Introduce polar coordinates in the $(\kappa_1,\kappa_2)$ plane, with
$\kappa=\sqrt{\kappa_1^2+\kappa_2^2}$ and $\theta$ the polar angle. 
This yields
\[
I= \int_{0}^{\infty}  \int_{0}^{2\pi} 
\frac{\kappa}{\kappa^2 + k^2} e^{-i\kappa( x \cos \theta+ y \sin \theta)}
\, d \theta \, d\kappa.
\]
Now, consider the integral over $\theta$ first. 
Introduce the notation
\[
\x=(x,y)=\rho (\cos \alpha,\sin\alpha), 
\]
for some $\alpha$. 
Then we have that 
\[
\x \cdot \kappab=(x,y)\cdot (\kappa_1,\kappa_2)=\kappa \rho (\cos\alpha \cos \theta + \sin\alpha \sin \theta)
= \kappa \rho \cos(\theta - \alpha), 
\]
and it follows that the inner integral is
\[
I_{\theta}=  \int_{0}^{2\pi}  e^{-i \kappa ( x \cos \theta+ y \sin \theta)} \, d \theta =
\int_{0}^{2\pi}  e^{-i \kappa  \rho \cos (\theta-\alpha)} \, d \theta =
\int_{0}^{2\pi}  e^{-i \kappa \rho \cos \theta} \, d \theta =
2 \int_{0}^{\pi}  e^{-i \kappa \rho \cos \theta} \, d \theta,
\]
where the second to last identity follows from the fact that we are integrating over a full period. 
From \cite{Gradshteyn2007}  p 912, 8.41, formula 7 for $\nu=0$ ($\Gamma(1/2)=\sqrt{\pi}$), 
\[
J_0(r \rho)=\frac{1}{\pi}\int_{0}^{\pi}  e^{-i r \rho \cos \theta} \, d \theta. 
\]
With this, the full integral becomes:
\[
 \bar{I}(k,x,y)=  2 \pi \int_{0}^{\infty} 
\frac{\kappa}{\kappa^2 + k^2} J_0( \kappa \rho) \, d\kappa.
\]
From \cite{Gradshteyn2007}, p 671, formula 4, 
\[
\int_{0}^{\infty}  \frac{r}{r^2 + k^2} J_0( ar) \, dr = K_0(ak), \quad a>0 \quad \Re(k) >0, 
\]
where $K_0$ is a modified Bessel function of the second kind. 
This yields
\[
 \bar{I}(k,x,y)=  2 \pi K_0(\rho k), 
\]
where $k$ is the positive square root of $k^2$. 

Let us now consider
\begin{align*}
I(k,x,y,\xi)=e^{-k^2/4\xi^2} \int_{\R^2}  \frac{1}{\kappa_1^2+\kappa_2^2+ k^2} 
e^{-(\kappa_1^2+\kappa_2^2)/4\xi^2} e^{-i\kappab \cdot \r}  \, d \kappa_1 d\kappa_2.
\end{align*} 
Following the same steps as above, this yields 
\begin{align*}
I(k,x,y,\xi)=
 2 \pi e^{-k^2/4\xi^2} \int_{0}^{\infty} 
\frac{\kappa}{\kappa^2 + k^2} J_0( \kappa \rho) \,
e^{-\kappa^2/4\xi^2} \, d\kappa.
\end{align*}
Unfortunately, we have not been able to directly find any closed expression for this
integral. 

We will instead use a technique suggested in Appendix C in \cite{Fripiat2010} to evaluate
$I(k,x,y,\xi)$. 
Before we do so, let us state the following result
\begin{equation}
\int_{-\infty}^{\infty} e^{-az^2} \, e^{-ipz} \, dz =
\sqrt{\frac{\pi}{a}} e^{-p^2/4a}, \quad \mbox{for  $a$ such that $\Re(a)>0$.}
\label{eqn:int_osc_exp}
\end{equation}

Now, denote $1/4\xi^2=\lambda$, and let 
$\tilde{I}(k,x,y,\lambda)=I(k,x,y,\xi)$. 
We will now evaluate $d\tilde{I}/d\lambda$. The goal
is to achieve a closed expression, that can then be integrated with
respect to $\lambda$ to achieve our final result. We get
\begin{align*}
\frac{d\tilde{I}}{d\lambda} & = 
- e^{-\lambda k^2} \int_{\R} \int_{\R}  
e^{-\lambda(\kappa_1^2+\kappa_2^2)} e^{-i(\kappa_1 x + \kappa_2 y)}
\, d \kappa_1 d\kappa_2 \  \\
& =
- e^{-\lambda k^2} 
\left[  \int_{\R} \left[  \int_{\R} 
e^{-i\kappa_1 x}    \, e^{-\lambda\kappa_1^2} \, d\kappa_1 
\right] 
e^{-i\kappa_2 y}    \, e^{-\lambda\kappa_2^2} \, d\kappa_2 
\right]=- \frac{\pi}{\lambda}  e^{-\lambda k^2}  e^{-(x^2+y^2)/4\lambda},  
\end{align*}
where we have used (\ref{eqn:int_osc_exp}) first for the integral over $\kappa_1$,
and then again for the integral over $\kappa_2$ in the last step. 
Considering that $\lim_{\lambda \rightarrow \infty} \tilde{I}(k,x,y,\lambda)=0$, we can
write (using $\rho^2=x^2+y^2$), 
\[
\tilde{I}(k,x,y,\lambda)=
\pi \int_{\lambda}^{\infty} \frac{1}{\alpha}  e^{-\alpha k^2}  e^{-\rho^2/4\alpha}\, d \alpha = 
\pi \int_{1}^{\infty} \frac{1}{t}  e^{-\lambda k^2 t}
e^{-\rho^2/(4\lambda t)}\, dt, 
\]
where we made a change of variables $\alpha=\lambda t$. 

Reintroducing $\lambda=1/4\xi^2$, we have 
\[
I(k,x,y,\xi)=
\pi \int_{1}^{\infty} \frac{1}{t}  e^{-\frac{k^2}{4\xi^2} t}
e^{-\rho^2\xi^2/t}\, dt = \pi K_0( \frac{k^2}{4\xi^2},\rho^2\xi^2), 
\]
using the definition given in (\ref{eqn:K0_two_args}).

\section{Asymptotic behavior of logarithmic sum.}
\label{app:logsum}

In this section, we will consider the sum of logarithmic terms 
that appears in (\ref{eqn:phiF_k0_1P}),
\[
S(x,y)=\sum_{n=1}^N q_n \log(\rho_n^2),
\]
where $\rho_n=|\r-\r_n|=\sqrt{(x-x_n)^2+(y-y_n)^2}$, 
under the assumption of charge neutrality, i.e.  that 
$\sum_{n=1}^N q_n=0$. 
First, let us establish the fact that
\begin{equation}
\sum_{n=1}^N q_n \log(\xi^2\rho_n^2)= \sum_{n=1}^N q_n \log(\rho_n^2),
\label{eqn:log_xi_rho}
\end{equation}
for $\xi$ constant. 
Using the laws of logarithms, we have
\begin{align*}
\sum_{n=1}^N q_n \log(\xi^2\rho_n^2)  =\sum_{n=1}^N q_n \left(
  \log(\xi^2)+ \log(\rho_n^2) \right) 
= \log(\xi^2) \sum_{n=1}^N q_n +\sum_{n=1}^N q_n
\log(\rho_n^2), 
\end{align*}
and using charge neutrality we obtain (\ref{eqn:log_xi_rho}). 

Consider the following expansion, valid for $0<z \le 2a$, 
\[
\log(z)=\log(a) + \sum_{p=1}^{\infty} (-1)^{p+1}\frac{(z-a)^p}{pa^p}.
\]
Now, use this formula to expand
$\log(\rho_n^2)=\log((x-x_n)^2+(y-y_n)^2)$ around $x^2+y^2$. 
This yields
\begin{align*}
\sum_{n=1}^N q_n \log(\rho_n^2) =
\sum_{n=1}^N q_n
\left[ \log(x^2+y^2)-
\sum_{p=1}^{\infty}
\frac{(-1)^{p+1}}{p}\frac{(-2xx_n-2yy_n+x_n^2+y_n^2))^p}{(x^2+y^2)^p}
\right] .
\end{align*}
The first term vanishes due to charge neutrality. Reordering and
explicitly writing out the leading order terms we get
\begin{align*}
\sum_{n=1}^N q_n \log(\rho_n^2) & =
- \frac{2}{x^2+y^2}\left[x \sum_{n=1}^N q_n x_n+y \sum_{n=1}^N q_n y_n \right]
+\frac{1}{x^2+y^2}  \sum_{n=1}^N q_n (x_n^2+y_n^2) \\
& -\frac{2}{(x^2+y^2)^2} \left[x^2 \sum_{n=1}^N q_n x_n^2+y^2
  \sum_{n=1}^N q_n y_n^2+2xy \sum_{n=1}^N q_n
  x_ny_n\right]+O(\frac{1}{r^3}), 
\end{align*}
with $r=\sqrt{x^2+y^2}$.


\bibliography{EwaldDerivations}{}

\begin{thebibliography}{10}

\bibitem{Abramowitz1970}
M.~Abramowitz and I.~Stegun.
\newblock {\em {Handbook of Mathematical Functions with Formulas, Graphs, and
  Mathematical Tables}}.
\newblock Dover, 1970.

\bibitem{Bertaut1952}
F.~Bertaut.
\newblock {L'\'{e}nergie \'{e}lectrostatique de r\'{e}seaux ioniques}.
\newblock {\em J. Phys. Radium}, 13:499, 1952.

\bibitem{Leeuw1979}
S.~W. de~Leeuw and J.~W. Perram.
\newblock {Electrostatic lattice sums for semi-infinite lattices}.
\newblock {\em Mol. Phys.}, {37}:{1313--1322}, 1979.

\bibitem{Leeuw1980}
S.~W. de~Leeuw, J.~W. Perram, and E.~R. Smith.
\newblock Simulation of electrostatic systems in periodic boundary conditions.
  i. lattice sums and dielectric constants.
\newblock {\em Proc. Royal Soc. London A}, 373:{27--56}, 1980.

\bibitem{Deserno1998}
M.~Deserno and C.~Holm.
\newblock {How to mesh up Ewald sums. I. A theoretical and numerical comparison
  of various particle mesh routines}.
\newblock {\em J. Chem. Phys.}, {109}:{7678--7693}, 1998.

\bibitem{Ewald1921}
P.~Ewald.
\newblock Die berechnung optischer und elektrostatischer gitterpotentiale.
\newblock {\em Ann. Phys.}, 64:{253--287}, 1921.

\bibitem{Fripiat2010}
J.G. Fripiat, J.~Delhalle, I.~Flamant, and F.~E. Harris.
\newblock {Ewald-type formulas for Gaussian-basis Bloch states in
  one-dimensionally periodic systems}.
\newblock {\em J. Chem. Phys.}, 132:044108, 2010.

\bibitem{Genovese2007}
L~Genovese, T~Deutsch, and S~Goedecker.
\newblock {Efficient and accurate three-dimensional Poisson solver for surface
  problems}.
\newblock {\em J. Chem. Phys.}, 127:054704, 2007.

\bibitem{Gradshteyn2007}
I.S. Gradshteyn and I.M. Ryzhik.
\newblock {\em {Table of Integrals, Series, and Products, Seventh Edition}}.
\newblock Academic Press, 2007.

\bibitem{Grzybowski2002}
A~Grzybowski and A~Br{\'o}dka.
\newblock {Electrostatic interactions in molecular dynamics simulation of a
  three-dimensional system with periodicity in one direction}.
\newblock {\em Molecular Physics}, 100(5):635--639, 2002.

\bibitem{Grzybowski2000}
A.~Grzybowski, E.~Gwozdz, and A.~Brodka.
\newblock {Ewald summation of electrostatic interactions in molecular dynamics
  of a three-dimensional system with periodicity in two directions}.
\newblock {\em Phys. Rev. B}, {61}:{6706--6712}, 2000.

\bibitem{Lindbo2011}
D.~Lindbo and A.-K. Tornberg.
\newblock {Spectral accuracy in fast Ewald-based methods for particle
  simulations}.
\newblock {\em J. Comput. Phys.}, 230(24):8744--8761, 2011.

\bibitem{Lindbo2012}
D.~Lindbo and A.-K. Tornberg.
\newblock {Fast and spectrally accurate Ewald summation for 2-periodic
  electrostatic systems}.
\newblock {\em J. Chem. Phys.}, 136(16):164111, 2012.

\bibitem{Parry1975}
D.~E. Parry.
\newblock The electrostatic potential in the surface region of an ionic
  crystal.
\newblock {\em Surf. Sci.}, 49:{433--440}, 1975.

\bibitem{Parry1976}
D.~E. Parry.
\newblock Errata; {The} electrostatic potential in the surface region of an
  ionic crystal.
\newblock {\em Surf. Sci.}, 54:{195--195}, 1976.

\bibitem{Porto2000}
M.~Porto.
\newblock {Ewald summation of electrostatic interactions of systems with finite
  extent in two of three dimensions}.
\newblock {\em J. Phys. A: Math. Gen.}, 2000.

\end{thebibliography}

\end{document}